\documentclass[twocolumn,english]{revtex4-1}
\usepackage[T1]{fontenc}
\usepackage[utf8]{luainputenc}
\usepackage{amsmath}
\usepackage{amssymb}
\usepackage{graphicx}
\usepackage{esint}
\usepackage{babel}
\begin{document}

\title{Generalized Coherent States, Reproducing Kernels, and Quantum Support
Vector Machines}

\author{Rupak Chatterjee$^{1,2}$}
\email{Rupak.Chatterjee@stevens.edu}

\author{Ting Yu$^{1,3}$}
\email{Ting.Yu@stevens.edu}

\affiliation{$^1$Center for Distributed Quantum Computing, Stevens Institute
of Technology, Castle Point on the Hudson, Hoboken, NJ 07030\\
$^2$Division of Financial Engineering, Stevens Institute of Technology, Castle Point on the Hudson, Hoboken, NJ 07030\\
$^3$Department of Physics and Engineering Physics, Stevens Institute of Technology, Castle Point on the Hudson, Hoboken, NJ 07030}








\begin{abstract}
The support vector machine (SVM) is a popular machine learning classification
method which produces a nonlinear decision boundary in a feature space
by constructing linear boundaries in a transformed Hilbert space.
It is well known that these algorithms when executed on a classical
computer do not scale well with the size of the feature space both
in terms of data points and dimensionality. One of the most significant
limitations of classical algorithms using non-linear kernels is that
the kernel function has to be evaluated for all pairs of input feature
vectors which themselves may be of substantially high dimension. This
can lead to computationally excessive times during training and during
the prediction process for a new data point. Here, we propose using
both canonical and generalized coherent states to rapidly calculate
specific nonlinear kernel functions. The key link will be the reproducing
kernel Hilbert space (RKHS) property for SVMs that naturally arise
from canonical and generalized coherent states. Specifically, we discuss
the fast evaluation of radial kernels through a positive operator
valued measure (POVM) on a quantum optical system based on canonical
coherent states. A similar procedure may also lead to fast calculations
of kernels not usually used in classical algorithms such as those
arising from generalized coherent states. 
\end{abstract}
\maketitle

\section{Introduction}

Machine learning is an area of mathematical statistics that uses computer
driven statistical learning techniques to find patterns in known empirical
data with the intent of applying these\emph{ learned} patterns on
new data. When one analyzes input data with the goal of predicting
or estimating a specific output, this is called \emph{supervised learning}.
An example would be using a financial firm's accounting reports to
determine it's credit rating. A machine learning algorithm would be
trained on the historical financial data of those firms with known
credit ratings and then be used either to predict a new rating for
a firm as new data arrives or to predict the rating of a firm with
no past credit rating history. Typically, one has $P$-features along
with $N$-observation\emph{ inputs }of these $P$-dimensional feature
vectors and a response called the \emph{supervising output} which
is measured on the same $N$-observation inputs. When a new set of
observations is given, one would like to predict or estimate the specific
output. A support vector machine (SVM) \cite{hastie2009elements,haykin2009neural}
is a supervised learning method that has been applied successfully
to a very wide range of binary classification problems (the supervised
output is binary) such as text classification (noun or verb), medical
risk (heart disease, no heart disease), homeland security (potential
risk, not a risk), etc. 

All machine learning models must necessarily deal with big data issues
such as the storage and fast retrieval of large amounts (\textasciitilde{}
petabytes) of data which becomes increasingly difficult as the feature
space and number of training observations of these features grow exponentially.\emph{
}Furthermore, many popular machine learning techniques use a multi-layered
or non-linear approach that leads to highly complex calculations resulting
in excessive runtime speeds. It will be shown that quantum support
vector machines based on coherent states may begin to address these
issues. The tensor product of coherent states allows an efficient
representation of high dimensional feature spaces. Quantum state overlap
measurements allow for the calculation of various non-linear SVM kernel
functions indicating a substantial runtime improvement over classical
algorithms.

A quantum version of a support vector classifier was given in \cite{rebentrost2014quantum}
where the authors provided a qubit representation of feature space
that was adaptable for simple polynomial type kernels. Yet, popular
nonlinear kernels such as those of the exponential or hyperbolic tangent
types are not easily amenable to their qubit representation. Here,
we propose using both canonical and generalized coherent states to
rapidly calculate these nonlinear kernel functions on high dimensional
feature spaces. A recent review of quantum machine learning techniques may be found in
\cite{biamonte2016quantum}.

\section{Support Vector Machines and Kernel Methods}

A short review of SVMs and a recent quantum version of a linear SVM
are described below. For more details, see \cite{bishop2006bishop,hastie2009elements,haykin2009neural,rebentrost2014quantum}.

Consider $N$ observations of a $P$-dimensional feature vector $\mathbf{X^{\mathit{i}}=(}X_{1}^{i},\;X_{2}^{i},\;X_{3}^{i}\ldots,\;X_{P}^{i})$.
Suppose that these observations live in a $P$-dimensional vector
space and are separable into two classes by a $P-1$ dimensional hyperplane.
For a 2-dimensional feature space, this hyperplane is simply a line
as the solid black line in figure 1 whereas in 3 dimensions, this
hyperplane will be a flat 2-dimensional plane.

\medskip{}

\begin{figure}
\center\includegraphics[scale=0.25]{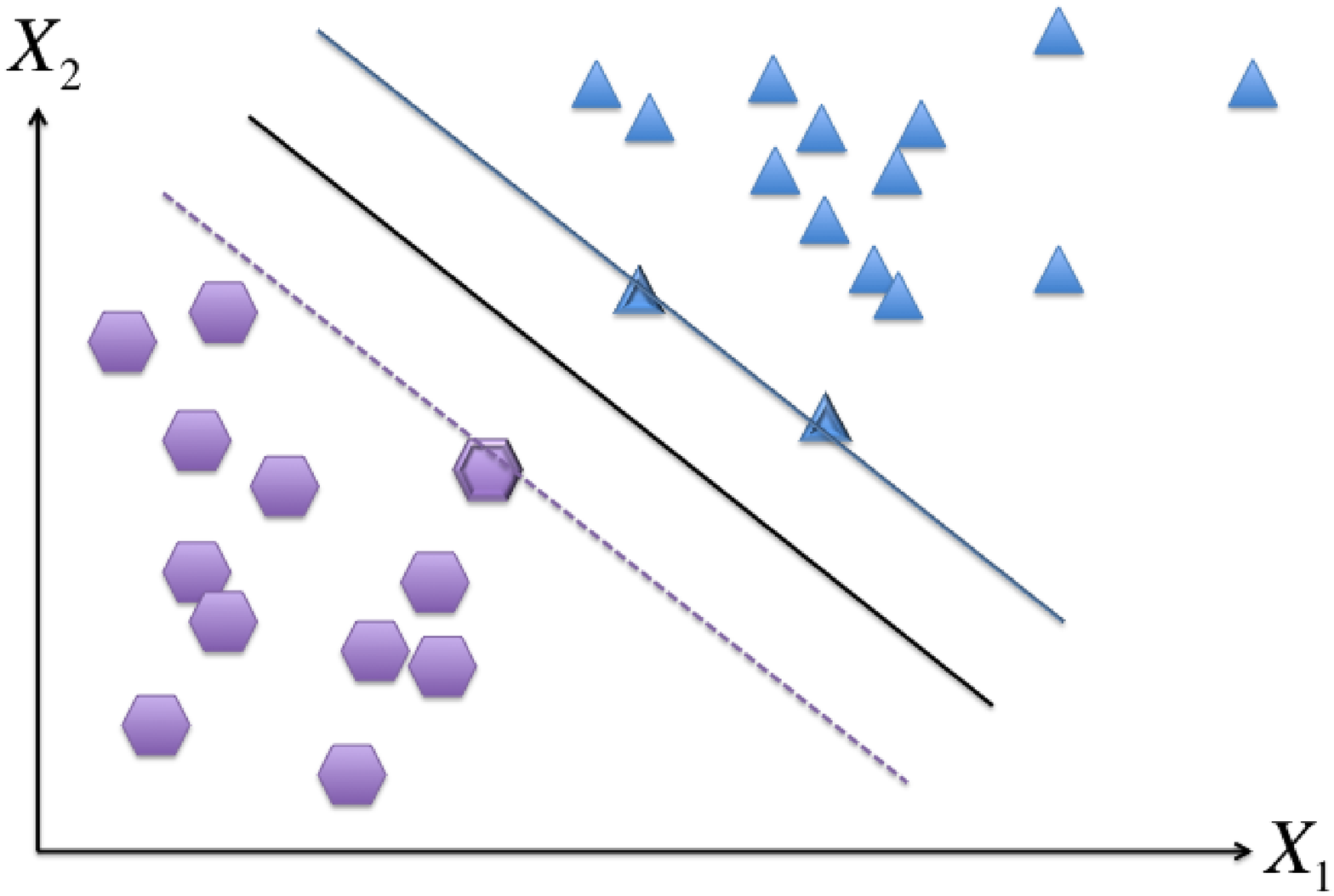}

\caption{Optimal Margin Classifier}
\end{figure}

Consider a $P$-dimensional weight vector $\mathbf{W=(}W_{1},\;W_{2},\ldots,\;W_{P})$
and a bias parameter $b$. A $P-1$ dimensional hyperplane is defined
by the equation

\begin{equation}
\mathbf{W\cdot X}+b=0.
\end{equation}
where $\mathbf{W}$ is the normal vector to the hyperplane. Since
our observations were assumed to be separable into two classes, each
observation $\mathbf{X}^{\mathit{i}}$ satisfies either

\begin{equation}
\mathbf{W\cdot X^{\mathrm{\mathit{i}}}}+b\leq0
\end{equation}
or
\begin{equation}
\mathbf{W\cdot X^{\mathrm{\mathit{i}}}}+b>0.
\end{equation}
Defining the supervised output as

\begin{equation}
y^{i}=\mathbf{SIGN}\left[\mathbf{W\cdot X^{\mathrm{\mathit{i}}}}+b\right],
\end{equation}
the two classes can be combined as 

\begin{equation}
y^{i}\left[\mathbf{W\cdot X^{\mathrm{\mathit{i}}}}+b\right]\geq0.
\end{equation}
The goal here is to find the optimal margin hyperplane (the best choice
of $\mathbf{W}$ and $b$) based on the $N$ training observations
$\mathbf{X}^{i}$ such that when a new test data point $\mathbf{X}^{*}$
needs to be analyzed, it will be correctly classified. This optimal
margin hyperplane will then act as a decision boundary for all new
data points. 

The perpendicular distance $M^{i}$ from the separating hyperplane
to a particular training point $\mathbf{X}^{i}$ is given by 

\begin{equation}
y^{i}\left[\mathbf{W\cdot X^{\mathrm{\mathit{i}}}}+b\right]=M^{i}
\end{equation}
Each training point will have a specific margin distance $M^{i}$
for a given set of hyperplane parameters $\mathbf{W}$ and $b$. The
optimality problem is to find the optimal hyperplane parameters that
gives the single biggest margin distance $M$ for all training points
simultaneously. In figure 1, the margin distance $M$ is the distance
between the solid black line, the decision boundary $f(\mathbf{X})=\mathbf{W\cdot X}+b=0$,
and the other two parallel lines that define the maximal margin region
of width $2M$. It is often the case that this problem has no perfect
solution. Therefore, rather than searching for a perfect decision
boundary, one can look for a hyperplane that separates \emph{most}
of the training data where a \emph{few feature vectors fall in the
margin region or on the wrong side of the decision boundary}. These
few observations are called \emph{support vectors }and are assigned
error terms (or slack variables) $\epsilon_{i}$ that are used to
violate the margin width $M$. The sum of these error terms are bounded,
i.e. $\sum_{i}\epsilon_{i}\leq K$.

As $\mathbf{W}$ is a normal vector to the hyperplane, one has $M=\left|\mathbf{W}\right|^{-1}$
. Therefore, maximizing $M$ is the same as minimizing the norm of
$\mathbf{W}$, 

\begin{equation}
Min_{\mathbf{W},b,\epsilon}\left|\mathbf{W}\right|\ s.t.\begin{cases}
y^{i}\left[\mathbf{W\cdot X^{\mathrm{\mathit{i}}}}+b\right]\geq(1-\epsilon_{i}) & ,\forall i\\
\\
\sum_{i=1}^{N}\epsilon_{i}\leq K & \epsilon_{i}\geq0
\end{cases}
\end{equation}
which may be rewritten as
\begin{equation}
\begin{array}{c}
Min_{\mathbf{W},b,\epsilon}\thinspace\thinspace\left[\dfrac{1}{2}\left|\mathbf{W}\right|^{2}+C\sum_{i=1}^{N}\epsilon_{i}\right]\\
\textrm{subject to}\\
y^{i}\left[\mathbf{W\cdot X^{\mathrm{\mathit{i}}}}+b\right]\geq(1-\epsilon_{i}),\epsilon_{i}\geq0,\forall i
\end{array}
\end{equation}
where $C$ controls the effect of the error terms coming from the
support vectors. If $C$ is very high, very few errors will be accepted
by the optimizer. $C=\infty$ reduces to the completely separable
case. 

One may use a Lagrange multiplier method to solve this problem. The
Lagrangian is given by
\begin{equation}
\begin{array}{c}
\mathfrak{\mathcal{L}}=\dfrac{1}{2}\left|\mathbf{W}\right|^{2}+C\sum_{i=1}^{N}\epsilon_{i}\\
-\sum_{i=1}^{N}\alpha^{i}\left\{ y^{i}\left[\mathbf{W\cdot X^{\mathrm{\mathit{i}}}}+b\right]-(1-\epsilon_{i})\right\} -\sum_{i=1}^{N}\mu^{i}\epsilon_{i}
\end{array}
\end{equation}
 with the optimality conditions given by the following minimizations,
\begin{equation}
\begin{array}{c}
\dfrac{\partial\mathcal{L}}{\partial\mathbf{W}}=0,\\
\\
\dfrac{\partial\mathcal{L}}{\partial b}=0,\\
\\
\dfrac{\partial\mathcal{L}}{\partial\epsilon_{i}}=0.\\
\\
\end{array}
\end{equation}
 The respective optimality conditions are 
\begin{equation}
\begin{array}{c}
\mathbf{W}=\sum_{i=1}^{N}\alpha^{i}y^{i}\mathbf{X}^{i},\\
\\
\sum_{i=1}^{N}\alpha^{i}y^{i}=0,\\
\\
\alpha^{i}=C-\mu^{i},\forall i.
\end{array}
\end{equation}
Note that the Lagrange multipliers must be positive, i.e. $\alpha^{i},\mu^{i}\geq0$
. Furthermore, only those Lagrange multipliers that exactly satisfy
\begin{equation}
\begin{array}{c}
\alpha^{i}\left\{ y^{i}\left[\mathbf{W\cdot X^{\mathrm{\mathit{i}}}}+b\right]-(1-\epsilon_{i})\right\} =0\\
\\
\mu^{i}\epsilon_{i}=0
\end{array}
\end{equation}
can have strictly nonzero values (the Krush-Kuhn-Tucker conditions,
see \cite{haykin2009neural}).

By substituting the solutions (11) into the Lagrangian (9), one obtains
the \emph{dual Lagrangian}

\begin{equation}
\mathfrak{\mathcal{L_{D}}}=\sum_{i=1}^{N}\alpha_{i}-\dfrac{1}{2}\sum_{i=1}^{N}\sum_{j=1}^{N}\alpha^{i}\alpha^{j}y^{i}y^{j}\mathbf{X^{\mathit{i}}\cdot X^{\mathit{j}}}
\end{equation}
Maximizing the dual Lagrangian with constraints $\sum_{i=1}^{N}\alpha^{i}y^{i}=0$
and $0\leq\alpha_{i}\leq C$ is often an easier problem than the minimization
(8) given above. In the dual problem, the Lagrange multipliers $\alpha^{i}$
are solved for and provide the optimal solution via the relation $\mathbf{W}=\sum_{i=1}^{N}\alpha^{i}y^{i}\mathbf{X}^{i}$.
The solution for the optimal hyperplane that solves the binary classification
problem is 
\begin{equation}
f(\mathbf{X})=\mathbf{W\cdot X}+b=\sum_{i=1}^{N}\alpha^{i}y^{i}\mathbf{X}^{i}\cdot\mathbf{X}+b
\end{equation}
It is useful to introduce the concept of a \emph{kernel }that can
be used to link support vector classifiers to the SVM technique below.
A kernel $K(\mathbf{X}^{\mathit{i}},\mathbf{X}^{j})$ is a type of
similarity measure between two observations and in the simple linear
case described here, it is given by
\begin{equation}
K(\mathbf{X}^{\mathit{i}},\mathbf{X}^{j})=\mathbf{X^{\mathit{i}}\cdot X^{\mathit{j}}}.
\end{equation}
This polynomial type kernel may be seen as a square symmetric matrix
with components given by (15). The dual Lagrangian and the optimal
hyperplane may be written as

\begin{equation}
\mathfrak{\mathcal{L_{D}}}=\sum_{i=1}^{N}\alpha_{i}-\dfrac{1}{2}\sum_{i=1}^{N}\sum_{j=1}^{N}\alpha^{i}y^{i}K(\mathbf{X^{\mathit{i}},X^{\mathit{j}})}\alpha^{j}y^{j}
\end{equation}
and

\begin{equation}
f(\mathbf{X})=\sum_{i=1}^{N}\alpha^{i}y^{i}K(\mathbf{X}^{i},\mathbf{X})+b.
\end{equation}

There are $N(N-1)/2$ dot products to calculate in (17) similar to
a square symmetric matrix where each dot products takes $O(P)$ time
to calculate. Finding the optimal $\alpha^{i}$ takes $O(N^{3})$
time. The convergence to an optimality error of $\epsilon$ is through
$O(log(1/\epsilon))$ iterations as shown in \cite{list2009svm}.
Therefore classically, the dual problem takes a computational time
of $O(log(1/\epsilon)N^{2}(P+N))$. This can be improved by a quantum
approach given in \cite{rebentrost2014quantum} briefly described
as follows.

Consider the following quantum representation of the observation feature
vector $\mathbf{X}^{\mathit{i}}$ using the base-2 \emph{bit string
configuration }introduced in \cite{rebentrost2014quantum} where $\left.|p\right\rangle =|p_{n-1}p_{n-2}p_{n-3}\cdots\,p_{2}p_{1}p_{0}\left.\right\rangle ,\;p=2^{0}p_{0}+2^{1}p_{1}+\cdots2^{n-1}p_{n-1},P=2^{n}$

\begin{equation}
\left.\mathrm{|}\mathbf{X}^{\mathit{i}}\right\rangle =\dfrac{1}{\left|\mathbf{X}^{i}\right|}\sum_{p=1}^{P}X_{p}^{(i)}\left.\mathrm{|}p\right\rangle .
\end{equation}
Using quantum parallelism, consider the following superposition state
of all the training data joined via a tensor product to an auxiliary
Hilbert space of computational basis states, $\left.\mathrm{|}\varPsi\right\rangle \in\mathcal{H}_{\mathit{\mathit{\mathcal{\mathit{a}}}ux}}\otimes\mathcal{H_{\mathit{trn}}}$,
where

\begin{equation}
\begin{array}{c}
\left.\mathrm{|}\varPsi\right\rangle =\dfrac{1}{\mathit{\mathcal{\sqrt{N_{\varPsi}}}}}\sum_{i=1}^{N}\left|\mathbf{X}^{i}\right|\left.\mathrm{|}i\right\rangle \otimes\left.\mathrm{|}\mathbf{X}^{\mathit{i}}\right\rangle ,\\
\\
\mathcal{N}_{\varPsi}=\sum_{i=1}^{N}\left|\mathbf{X}^{i}\right|^{2}.
\end{array}
\end{equation}
The density operator associated with this state is given by
\begin{equation}
\rho_{aux-trn}=\dfrac{1}{\mathit{\mathcal{N_{\varPsi}}}}\sum_{i=1}^{N}\sum_{j=1}^{N}\left|\mathbf{X}^{i}\right|\left|\mathbf{X}^{j}\right|\left.\mathrm{|}i\right\rangle \left\langle j|\right.\otimes\left.\mathrm{|}\mathbf{X}^{\mathit{i}}\right\rangle \left\langle \mathbf{X}^{\mathit{j}}|\right.
\end{equation}
Taking a partial trace over $\mathcal{H_{\mathit{trn}}}$ gives

\begin{equation}
\rho_{aux}=\dfrac{1}{\mathit{\mathcal{N_{\varPsi}}}}\sum_{i=1}^{N}\sum_{j=1}^{N}\left\langle \mathbf{X}^{\mathit{i}}\right.|\left.\mathbf{X}^{j}\right\rangle \left|\mathbf{X}^{i}\right|\left|\mathbf{X}^{j}\right|\left.\mathrm{|}i\right\rangle \left\langle j|\right.
\end{equation}
From equation (18), it is clear that

\begin{equation}
\mathbf{X}^{i}\cdot\mathbf{X}^{j}=\left\langle \mathbf{X}^{\mathit{i}}\left.\mathrm{|}\mathbf{X}^{\mathit{j}}\right\rangle \right.\left|\mathbf{X}^{i}\right|\left|\mathbf{X}^{j}\right|
\end{equation}
and therefore
\begin{equation}
\rho_{aux}=\dfrac{1}{\mathit{\mathcal{N_{\varPsi}}}}\sum_{i=1}^{N}\sum_{j=1}^{N}K(\mathbf{X}^{\mathit{i}},\mathbf{X}^{j})\left.\mathrm{|}i\right\rangle \left\langle j|\right.=\dfrac{K}{\mathbf{Tr}K}
\end{equation}
where
\begin{equation}
\mathbf{Tr}K=\sum_{i=1}^{N}K(\mathbf{X}^{\mathit{i}},\mathbf{X}^{i})=\sum_{i=1}^{N}\mathbf{X^{\mathit{i}}\cdot X^{\mathit{i}}}=\sum_{i=1}^{N}\left|\mathbf{X}^{i}\right|^{2}=\mathcal{N}_{\varPsi}
\end{equation}
According to \cite{rebentrost2014quantum}, by using this method to
calculate the kernel matrix and furthermore turning the optimization
problem into a quantum matrix inversion problem, the runtime for their
quantum support vector classifier method becomes \textasciitilde{}$O(log(PN))$.

Consider the classification problem illustrated in figure 2. Visually,
one sees a distinct boundary on the left side of the figure yet it
is clearly not linear. Neither of the previous two methods will find
a satisfactory decision hyperplane. However, a non-linear decision
boundary appears to be possible. Consider a set $\left\{ \varphi_{j}(\mathbf{X})\right\} _{j=1}^{\infty}$
of square integrable functions that map the finite dimensional feature
space into an infinite dimensional space. The equation in this new
space, analogous to (1), is given by
\begin{equation}
\sum_{j=1}^{\infty}W_{j}\varphi_{j}(\mathbf{X})=0
\end{equation}
with an optimal classification hyperplane of
\begin{equation}
f_{SVM}(\mathbf{X})=\sum_{i=1}^{N}\alpha^{i}y^{i}\sum_{j=1}^{\infty}\varphi_{j}(\mathbf{X}^{i})\varphi_{j}(\mathbf{X})+b
\end{equation}
(the right hand side of figure 2). One would like to define an inner
product kernel
\begin{equation}
K(\mathbf{X}^{i},\mathbf{X})=\sum_{j=1}^{\infty}\varphi_{j}(\mathbf{X}^{i})\varphi_{j}(\mathbf{X})
\end{equation}
such that (26) once again reduces to the form given in (17), 
\begin{equation}
f_{SVM}(\mathbf{X})=\sum_{i=1}^{N}\alpha^{i}y^{i}K(\mathbf{X}^{i},\mathbf{X})+b.
\end{equation}
Clearly, specifying the kernel is sufficient to find the optimal classification
boundary. These methods are often referred to as kernel methods \cite{bishop2006bishop}
as one does not need the explicit mapping $\varphi_{j}(\mathbf{X})$
itself as the kernel alone defines the solution in equation (28) (``the
kernel trick'' or more formally the Representer Theorem \cite{haykin2009neural,hofmann2008kernel}). 

The form of (27) is possible using Mercer's theorem from functional
analysis \cite{berlinet2011reproducing,courant1966methods} which
comes down to a spectral decomposition of a continuous symmetric kernel
using a eigenvalue-eigenfunction expansion (see section IV below),
\begin{equation}
K(\mathbf{X}^{'},\mathbf{X})=\sum_{j=1}^{\infty}\lambda_{j}\varphi_{j}(\mathbf{X}^{'})\varphi_{j}(\mathbf{X})
\end{equation}
where the special case of $\lambda_{i}=1,\forall i$ has been used
in (27). Some popular kernels in the SVM literature are

\emph{Polynomial of degree d:
\begin{equation}
K(\mathbf{X}^{\mathit{i}},\mathbf{X}^{j})=(1+\mathbf{X}^{\mathit{i}}\cdot\mathbf{X}^{j})^{d}
\end{equation}
}

\emph{Radial Kernels (Gaussian):
\begin{equation}
K\mathbf{(\mathbf{X}^{\mathit{i}},\mathbf{X}^{\mathit{j}})=exp}\left(-\dfrac{1}{2\sigma^{2}}\left|\mathbf{X}^{\mathit{i}}-\mathbf{X}^{j}\right|^{2}\right)
\end{equation}
}

\emph{Radial Kernels (Ornstein Uhlenbeck):}

\emph{
\begin{equation}
K(\mathbf{X}^{\mathit{i}},\mathbf{X}^{j})=\mathbf{exp}\left(-\dfrac{1}{\gamma}\left|\mathbf{X}^{\mathit{i}}-\mathbf{X}^{j}\right|\right)
\end{equation}
}

\emph{Sigmoidal (two-layer perceptron):
\begin{equation}
K\mathbf{(\mathbf{X}^{\mathit{i}},\mathbf{X}^{\mathit{j}})=tanh}[\alpha+\mathbf{\beta X}^{\mathit{i}}\cdot\mathbf{X}^{j}]
\end{equation}
}

\begin{figure}
\center\includegraphics[scale=0.45]{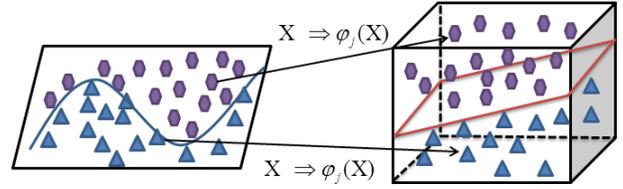}

\caption{Support Vector Machine}
\end{figure}
By using these types of kernels, one is looking for a hyperplane in
a \emph{higher dimensional feature space.} This\emph{ decision boundary}
hyperplane in the higher dimensional space, as in the right side of
figure 2, results in a \emph{nonlinear decision boundary} in the original
feature space.

Analogous to the evaluation issues of (17), one of the most significant
limitations of classical algorithms using non-linear kernels is that
the kernel function has to be evaluated for all pairs of input feature
vectors $\mathbf{X}^{\mathit{i}}$ and $\mathbf{X}^{j}$ which themselves
may be of substantially high dimension. This can lead to computationally
excessive times during training and during the prediction process
for new data points. In fact, classical methods such as sparse kernel
methods \cite{bishop2006bishop} have been developed to deal with
this issue but they are mostly heuristic methods used to increase
runtime speeds. Rather, quantum methods analogous to the linear kernel
methods described in (18)-(24) would be highly desirable and beneficial
to using nonlinear SVMs in a fast and efficient manner. We now demonstrate
how this is possible using coherent states.

\section{Radial Kernels from Canonical Coherent States}

A feature space representation using canonical coherent states may
be acheived as follows. Consider the $P$-dimensional tensor product
of canonical coherent states 
\begin{equation}
\left.|\alpha_{1}\right\rangle =\left.|\alpha_{1}^{1}\right\rangle \otimes\left.|\alpha_{1}^{2}\right\rangle \otimes\cdots\otimes\left.|\alpha_{1}^{p}\right\rangle ,
\end{equation}
\begin{equation}
\left.|\alpha_{2}\right\rangle =\left.|\alpha_{2}^{1}\right\rangle \otimes\left.|\alpha_{2}^{2}\right\rangle \otimes\cdots\otimes\left.|\alpha_{2}^{p}\right\rangle ,
\end{equation}
and their overlap 
\begin{equation}
\begin{array}{c}
\left|\left\langle \alpha_{1}\right.|\left.\alpha_{2}\right\rangle \right|^{2}=\\
\mathbf{exp}\left(-\left|\alpha_{1}^{1}-\alpha_{2}^{1}\right|^{2}\right)\mathbf{exp}\left(-\left|\alpha_{1}^{2}-\alpha_{2}^{2}\right|^{2}\right)\cdots
\end{array}
\end{equation}
In a feature space with $\mathbf{X}^{1}=(\alpha_{1}^{1},\alpha_{1}^{2},\ldots,\alpha_{1}^{p})$
and $\mathbf{X}^{2}=(\alpha_{2}^{1},\alpha_{2}^{2},\ldots,\alpha_{2}^{p})$,
this becomes
\begin{equation}
\left|\left\langle \alpha_{1}\right.|\left.\alpha_{2}\right\rangle \right|^{2}=\mathbf{exp}\left(-\left|\mathbf{X}^{\mathit{\mathbf{\mathrm{1}}}}-\mathbf{X}^{2}\right|^{2}\right)
\end{equation}
which is the popular radial kernel form of SVMs.

\begin{figure}
\center\includegraphics[scale=0.3]{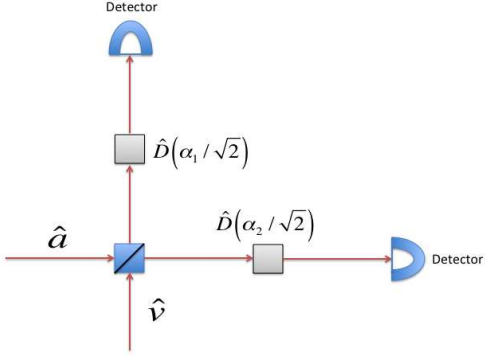}

\caption{Coherent State Overlap Measurement }
\end{figure}

Several simple measurements \cite{banaszek1999optimal,becerra2013implementation}
may produce this overlap function in computation times faster than
the time needed on a classical computer. Consider figure 3 which has
been adapted from \cite{banaszek1999optimal}. The annihilation operator
is related to one of two coherent states $\left.|\alpha_{1}\right\rangle $
or $\left.|\alpha_{2}\right\rangle $. This unknown state enters the
50-50 beam splitter along with a vacuum field whereupon the output
states undergo a coherent displacement given by
\begin{equation}
D(\alpha_{i}/\sqrt{2})=\mathbf{exp}\left[\frac{\alpha_{i}}{\sqrt{2}}\hat{a}^{\dagger}-\frac{\alpha_{i}^{*}}{\sqrt{2}}\hat{a}\right]
\end{equation}
This displacement operator acting on the vacuum state (indicated by
$\hat{v}$ in figure 3) creates two coherent states $\left.|\beta_{1}\right\rangle $
and $\left.|\beta_{2}\right\rangle $ that have operators $\hat{b}_{1}$
and $\hat{b}_{2}$ related to the incident fields via a Hadamard gate,
\begin{equation}
\left(\begin{array}{c}
\hat{b}_{1}\\
\hat{b}_{2}
\end{array}\right)=\frac{1}{\sqrt{2}}\left(\begin{array}{cc}
1 & 1\\
1 & -1
\end{array}\right)\left(\begin{array}{c}
\hat{a}\\
\hat{v}
\end{array}\right)
\end{equation}
For the joint measurement performed by the two detectors, one has
the following four outcomes with projections given by 
\begin{equation}
P_{1}=\left.|\beta_{1}\right\rangle \left\langle \beta_{1}|\right.\otimes\left(I-\left.|\beta_{2}\right\rangle \left\langle \beta_{2}|\right.\right),
\end{equation}
\begin{equation}
P_{2}=\left(I-\left.|\beta_{1}\right\rangle \left\langle \beta_{1}|\right.\right)\otimes\left.|\beta_{2}\right\rangle \left\langle \beta_{2}|\right.,
\end{equation}
\begin{equation}
P_{3}=\left.|\beta_{1}\right\rangle \left\langle \beta_{1}|\right.\otimes\left.|\beta_{2}\right\rangle \left\langle \beta_{2}|\right.,
\end{equation}
and
\begin{equation}
P_{4}=\left(I-\left.|\beta_{1}\right\rangle \left\langle \beta_{1}|\right.\right)\otimes\left(I-\left.|\beta_{2}\right\rangle \left\langle \beta_{2}|\right.\right),
\end{equation}
along with the corresponding POVM, 
\begin{equation}
\varPi_{i}=\left\langle 0|\right.P_{i}\left.|0\right\rangle ,\;i=1,...,4.
\end{equation}

Following \cite{banaszek1999optimal}, consider the following explicit
representation, using normal ordering, of the coherent state projection
operator,
\begin{equation}
\left.|\beta_{i}\right\rangle \left\langle \beta_{i}|\right.=:\mathbf{exp}\left[-\left(\hat{b}_{i}^{\dagger}-\frac{\alpha_{i}^{*}}{\sqrt{2}}\right)\left(\hat{b}_{i}-\frac{\alpha_{i}}{\sqrt{2}}\right)\right]:,\,i=1,2.
\end{equation}
Oone can show that $\left\langle 0|\right.\left.\beta_{1}\right\rangle \left\langle \beta_{1}\right.\left.|0\right\rangle $
is given by,
\begin{equation}
\begin{array}{c}
\left\langle 0|\right.:\mathbf{exp}\left[-\left(\hat{b}_{1}^{\dagger}-\frac{\alpha_{1}^{*}}{\sqrt{2}}\right)\left(\hat{b}_{1}-\frac{\alpha_{1}}{\sqrt{2}}\right)\right]:\left.|0\right\rangle =\\
\left\langle 0|\right.:\mathbf{exp}\left[-\left(\frac{\hat{a}^{\dagger}+\hat{v}}{\sqrt{2}}-\frac{\alpha_{1}^{*}}{\sqrt{2}}\right)\left(\frac{\hat{a}+\hat{v}}{\sqrt{2}}-\frac{\alpha_{1}}{\sqrt{2}}\right)\right]:\left.|0\right\rangle 
\end{array}
\end{equation}
which simplifies to
\begin{equation}
:\mathbf{exp}\left[-\frac{1}{2}\left(\hat{a}^{\dagger}-\alpha_{1}^{*}\right)\left(\hat{a}-\alpha_{1}\right)\right]:\doteq:\hat{R}_{1}:
\end{equation}
where we have defined the operator $\hat{R}_{i}=\mathbf{exp}\left[-\frac{1}{2}\left(\hat{a}^{\dagger}-\alpha_{i}^{*}\right)\left(\hat{a}-\alpha_{i}\right)\right]$.
Following in this manner, it can be verified that the POVM (44) may
now be written as in \cite{banaszek1999optimal},

\begin{equation}
\varPi_{1}=:\hat{R}_{1}:-:\hat{R}_{1}\hat{R}_{2}:,
\end{equation}

\begin{equation}
\varPi_{2}=:\hat{R}_{2}:-:\hat{R}_{1}\hat{R}_{2}:,
\end{equation}
\begin{equation}
\varPi_{3}=:\hat{R}_{1}\hat{R}_{2}:,
\end{equation}
and
\begin{equation}
\varPi_{4}=I-:\hat{R}_{1}:-:\hat{R}_{2}:+:\hat{R}_{1}\hat{R}_{2}:,
\end{equation}
Finally, the non zero probabilities of measuring one of the states
$\left.|\alpha_{1}\right\rangle $ or $\left.|\alpha_{2}\right\rangle $
are given by
\begin{equation}
\left\langle \alpha_{1}|\right.\varPi_{1}\left.|\alpha_{1}\right\rangle =1-\mathbf{exp}\left(-\left|\alpha_{1}-\alpha_{2}\right|^{2}/2\right),
\end{equation}
\begin{equation}
\left\langle \alpha_{2}|\right.\varPi_{2}\left.|\alpha_{2}\right\rangle =1-\mathbf{exp}\left(-\left|\alpha_{1}-\alpha_{2}\right|^{2}/2\right)
\end{equation}
and
\begin{equation}
\left\langle \alpha_{1}|\right.\varPi_{3}\left.|\alpha_{1}\right\rangle =\left\langle \alpha_{2}|\right.\varPi_{3}\left.|\alpha_{2}\right\rangle =\mathbf{exp}\left(-\left|\alpha_{1}-\alpha_{2}\right|^{2}/2\right)
\end{equation}
which provides the necessary evaluation of the radial kernel for our
quantum SVM.

How has this method improved the classical runtime? Classically, there
are $N(N-1)/2$ exponential pairs to evaluate where each exponential of a dot
product takes $O(P^{3})$ time to calculate. Finding the
optimal $\alpha^{i}$ takes $O(N^{3})$ time while the convergence
to an optimality error of $\epsilon$ is through $O(log(1/\epsilon))$
iterations \cite{list2009svm} leading to a total classical computational
time of $O(log(1/\epsilon)N^{2}(P^{3}+N))$. The POVM probabilities
(52)-(54) converge after $M$ repeated measurements. For the case
of $(P,N)\gg M$, which is common in the big data arena, the calculation
of the radial kernel via the POVM set-up becomes largely independent
of $P$ (due to the tensor product Hilbert space representation of
the $P$-dimensional feature vector in the quantum domain). The classical
time has been effectively reduced to $O(log(1/\epsilon)N^{2}(1+N))$.
For data sets where $P\sim N$, this is a substantial improvement.

The $O(N^{2})$ time coming from the $N(N-1)/2$ pairs of feature
vectors may be further reduced by generalizing this methodology to
multiple coherent states. The two state experiment above was generalized
to four coherent states in \cite{becerra2013implementation} where
the POVM to unambiguously identify one of the possible non-orthogonal
states (unambiguous state discrimination or USD) are

\begin{equation}
\begin{array}{c}
\varPi_{i}^{USD}=:\hat{R}_{i}:-\sum_{i<j\:\&\:cyclic}^{4}:\hat{R}_{i}\hat{R}_{j}:\\
+\sum_{i<j<k\:\&\:cyclic}^{4}:\hat{R}_{i}\hat{R}_{j}\hat{R}_{k}:\\
-:\hat{R}_{1}\hat{R}_{2}\hat{R}_{3}\hat{R}_{4}:,\;i=1...,4
\end{array}
\end{equation}
Note that this formula is only an extension of $\varPi_{1}$ and $\varPi_{2}$
above as neither $\varPi_{3}$ or $\varPi_{4}$ provides a USD result.
Therefore, the authors of \cite{becerra2013implementation} add an
inconclusive USD measurement $\varPi_{5}^{USD}=1-\sum_{i=1}^{4}\varPi_{i}^{USD}$
to complete the POVM set. (55) has a natural extension for an arbitrary
number of coherent states. From this generalization, we postulate
that the $O(N^{2})$ pair calculation time may be reduced by the parallel
measurement of several pairs of coherent states but a further study
will be needed to get an estimate of the reduction in runtime.

Finally, the $O(N^{3})$ time coming from finding the optimal $\alpha^{i}$
may also be reduced as follows. In \cite{suykens1999least}, a least
squares method is introduced to solve the quadratic programming problem
of SVMs. Rebentrost et al. \cite{rebentrost2014quantum} take this
least squares method and propose an approximate quantum least squares
method for their polynomial based kernel SVM using a quantum matrix
inversion algorithm. Their method is largely independent of the type
of kernel and therefore the quantum radial kernel calculation described
here may be combined with their method to substantially reduce the
$O(N^{3})$ runtime of the classical quadratic programming problem
(\cite{rebentrost2014quantum} reduce their polynomial kernel problem
to $\sim O(logN)$).

\section{Reproducing Kernel Hilbert Spaces and Mercer's Theorem}

As the structure of SVMs comes down to the kernel function, a more
precise description of its structure will lead naturally to a relation
with other types of coherent states rather than just the canonical
ones used above. For more details, see \cite{berlinet2011reproducing,haykin2009neural,hofmann2008kernel}.

Let $\mathcal{H}$ be a Hilbert space of real square integrable functions
$f(\mathbf{X})$ defined on a set $\mathcal{X}$. A \emph{Mercer Kernel}
is symmetric kernel $K:\mathcal{X}\times\mathcal{X}\rightarrow\mathbb{R}$,
$K(\mathbf{X}^{'},\mathbf{X})=K(\mathbf{X},\mathbf{X^{'}})$ with
the following positive semi-definite property,
\begin{equation}
\int\int f(\mathbf{X}^{'})K(\mathbf{X}^{'},\mathbf{X})f(\mathbf{X})d\mathbf{X}^{'}d\mathbf{X}\geq0,\forall f(\mathbf{X}).
\end{equation}

\emph{Mercer's Theorem: }For a Mercer kernel $K(\mathbf{X}^{'},\mathbf{X})$
, there exists a set of orthonormal functions $\varphi_{i}(\mathbf{X})$
\begin{equation}
\int\varphi_{i}(\mathbf{X})\varphi_{j}(\mathbf{X})d\mathbf{X}=\delta_{ij}
\end{equation}
such that 
\begin{equation}
K(\mathbf{X}^{'},\mathbf{X})=\sum_{j=1}^{\infty}\lambda_{j}\varphi_{j}(\mathbf{X}^{'})\varphi_{j}(\mathbf{X}),\ \lambda_{j}>0.
\end{equation}

\emph{Reproducing Kernel Hilbert Space} (RKHS): Let $\mathcal{H}$
be a Hilbert space of real functions $f$ defined on a set $\mathcal{X}$.
$\mathcal{H}$ is called a reproducing kernel Hilbert space with an
inner product $\left\langle \cdot\right.|\left.\cdot\right\rangle $
if there exists a kernel function $K:\mathcal{X}\times\mathcal{X}\rightarrow\mathbb{R}$
with the following properties, 
\begin{equation}
K(\mathbf{X}^{'},\mathbf{X})\in\mathcal{H},\forall\mathbf{X^{'}},
\end{equation}
(if $\mathbf{X}^{'}$ is an index then $K(\mathbf{X}^{'},\mathbf{X})$
may be seen as a function of $\mathbf{X}$) and the \emph{reproducing
property
\begin{equation}
\left\langle K(\cdot,\mathbf{X)}\right.|\left.f(\cdot)\right\rangle =f(\mathbf{X}).
\end{equation}
}Property (59) can be used with Mercer type kernels to induce a RKHS.
The complex version of these kernels are often referred to as Bergman
kernels \cite{ali2000coherent} which produce RKHS of square integrable
holomorphic functions. All the standard kernels used in SVMs satisfy
Mercer's theorem \cite{haykin2009neural} and may be used to create
RKHSs.

It is also possible to create new kernels from existing ones. Suppose
one has two kernels $K_{1}(\mathbf{X}^{'},\mathbf{X})$ and $K_{2}(\mathbf{X}^{'},\mathbf{X})$.
One can add and multiply these kernels to produce a new kernel $K_{3}(\mathbf{X}^{'},\mathbf{X})$,
i.e.
\begin{equation}
K_{3}(\mathbf{X}^{'},\mathbf{X})=K_{1}(\mathbf{X}^{'},\mathbf{X})+K_{2}(\mathbf{X}^{'},\mathbf{X}),
\end{equation}
or
\begin{equation}
K_{3}(\mathbf{X}^{'},\mathbf{X})=K_{1}(\mathbf{X}^{'},\mathbf{X})\cdot K_{2}(\mathbf{X}^{'},\mathbf{X}).
\end{equation}

\section{A View Towards Generalized Coherent States}

In order for the quantum SVM method to go beyond radial kernels, one
needs to analyze generalized coherent states and their relation to
RKHSs. The generalization of canonical coherent states has historically
proceeded along three (not necessarily equivalent) lines of thought
which happen to result in equivalent definitions for canonical states.
One generalization initiated by Barut and Girardello \cite{barut1971}
follows the path of creating generalized coherent states as eigenstates
of a specific operator from a Lie algebra. Another group theoretical
generalization begun independently by Perelomov \cite{perelomov1972coherent,perelomov2012generalized}
and Gilmore \cite{gilmore1972geometry} considers a generalized displacement
operator acting on a vacuum state. Here, we wish to consider a third
approach started by Klauder and Gazeau \cite{klauder1996coherent,gazeau1999coherent,gazeau2000generalized,ali2000coherent,gazeau2009coherent}
based on the following definition.

\emph{Definition: }Coherent states $\left.|\alpha\right\rangle $
are \emph{wave-packets} that are superpositions of eigenstates $\left.|n\right\rangle $
of a self-adjoint operator and square integrable functions $\psi_{n}(\alpha)$,
such that $\left.|\alpha\right\rangle =\sum_{n=0}^{\infty}\psi_{n}^{*}(\alpha)\left.|n\right\rangle $,
where the states are normalized $\left\langle \alpha\right.|\left.\alpha\right\rangle =1,$
with a resolution of identity given by $\int_{\mathbb{C}}\left.|\mathbf{\alpha}\right\rangle \left\langle \alpha|\right.d^{2}\alpha/\pi=I$.

This generalization has a natural structure of an underlying RKHS.
A resolution of identity requirement leads to the existence of a POVM.
Let $A$ be a self-adjoint operator with a discrete spectrum $\{a_{n}\}$
and normalized eigenstates $\left\{ \left.|a_{n}\right\rangle \right\} $
that form an orthonormal basis in a separable complex Hilbert space
$\mathcal{H}$,
\begin{equation}
A\left.|a_{n}\right\rangle =a_{n}\left.|a_{n}\right\rangle .
\end{equation}
Consider a measure space $(\mathcal{X},\mu)$ and the Hilbert space
$L^{2}(\mathcal{X},\mu)$ of all square-integrable functions $\psi(\mathbf{X})$
on the set $\mathcal{X}$ ($\mathbf{X}\in\mathcal{X}$),
\begin{equation}
\left\langle \psi(\mathbf{X)}\right.|\left.\psi(\mathbf{X})\right\rangle =\int_{\mathcal{X}}\left|\psi(\mathbf{X})\right|^{2}\mu(d\mathbf{X})<\infty
\end{equation}
with orthonormal basis functions $\left\{ \psi_{n}(\mathbf{X})\right\} $
\begin{equation}
\left\langle \psi_{n}(\mathbf{X)}\right.|\left.\psi_{m}(\mathbf{X})\right\rangle =\int_{\mathcal{X}}\psi_{n}^{*}(\mathbf{X})\psi_{m}(\mathbf{X})\mu(d\mathbf{X})=\delta_{nm}.
\end{equation}
Furthermore, assume that the eigenstates $\left\{ \left.|a_{n}\right\rangle \right\} $
are in a one-to-one correspondence with these orthonormal basis functions.
Generalized coherent states may be defined by \cite{ali2000coherent,gazeau2009coherent}
\begin{equation}
\left.|\mathbf{X}\right\rangle =\dfrac{1}{\sqrt{\mathcal{N}(\mathbf{X})}}\sum_{n=0}^{\infty}\psi_{n}^{*}(\mathbf{X})\left.|a_{n}\right\rangle ,
\end{equation}
with a normalization restriction of 
\begin{equation}
0<\mathcal{N}(\mathbf{X})=\sum_{n=0}^{\infty}\left|\psi_{n}(\mathbf{X})\right|^{2}\leq\infty.
\end{equation}
The resolution of identity in $\mathcal{H}$ is
\begin{equation}
\int_{\mathcal{X}}\left.|\mathbf{X}\right\rangle \left\langle \mathbf{X}|\right.\mathcal{N}(\mathbf{X})\mu(d\mathbf{X})=I
\end{equation}

For any $\left.|\psi\right\rangle \in\mathcal{H}$, one can associate
an element $\psi(\mathbf{X})\in L^{2}(\mathcal{X},\mu)$ as
\begin{equation}
\psi(\mathbf{X})=\sqrt{\mathcal{N}(\mathbf{X})}\left\langle \mathbf{X}\right.|\left.\psi\right\rangle 
\end{equation}
Note that there is a natural isomorphism between the Hilbert space
$\mathcal{H}$ and $L^{2}(\mathcal{X},\mu)$ because of the one to
one correspondence of basis functions explicitly given by the special
case of (69),
\begin{equation}
\psi_{n}(\mathbf{X})=\sqrt{\mathcal{N}(\mathbf{X})}\left\langle \mathbf{X}\right.|\left.a_{n}\right\rangle .
\end{equation}
Using the resolution of identity (68) in (69), one has
\begin{equation}
\psi(\mathbf{X})=\sqrt{\mathcal{N}(\mathbf{X})}\int_{\mathcal{X}}\left\langle \mathbf{X}\right.|\left.\mathbf{X}^{\prime}\right\rangle \left\langle \mathbf{X}^{\prime}\right.|\left.\psi\right\rangle \mathcal{N}(\mathbf{X}^{\prime})\mu(d\mathbf{X}^{\prime})
\end{equation}
which may be written as
\begin{equation}
\psi(\mathbf{X})=\int_{\mathcal{X}}\sqrt{\mathcal{N}(\mathbf{X})}\left\langle \mathbf{X}\right.|\left.\mathbf{X}^{\prime}\right\rangle \dfrac{\psi(\mathbf{X}^{\prime})}{\sqrt{\mathcal{N}(\mathbf{X}^{\prime})}}\mathcal{N}(\mathbf{X}^{\prime})\mu(d\mathbf{X}^{\prime})
\end{equation}
or
\begin{equation}
\psi(\mathbf{X})=\int_{\mathcal{X}}\sqrt{\mathcal{N}(\mathbf{X})\mathcal{N}(\mathbf{X}^{\prime})}\left\langle \mathbf{X}\right.|\left.\mathbf{X}^{\prime}\right\rangle \psi(\mathbf{X}^{\prime})\mu(d\mathbf{X}^{\prime}).
\end{equation}
One may create a RKHS $\mathcal{H_{K}}$ composed of elements given
by (69) spanned by basis functions (70). By defining a kernel as
\begin{equation}
K(\mathbf{\mathbf{X}{}^{\prime}},\mathbf{X})=\sqrt{\mathcal{N}(\mathbf{X})\mathcal{N}(\mathbf{X}^{\prime})}\left\langle \mathbf{X}\right.|\left.\mathbf{X}^{\prime}\right\rangle ,
\end{equation}
equation (73) becomes identical to the kernel reproducing property
(60),

\begin{equation}
\psi(\mathbf{X})=\int_{\mathcal{X}}K(\mathbf{\mathbf{X}{}^{\prime}},\mathbf{X})\psi(\mathbf{X}{}^{\prime})\mu(d\mathbf{X}^{\prime})=\left\langle K(\cdot,\mathbf{X})\right.|\left.\psi(\cdot)\right\rangle 
\end{equation}
making $\mathcal{H_{K}}$ a RKHS. For standard canonical coherent
states, one has $\mathcal{X}=\mathbb{C}$, $\mathcal{N}(\mathbf{X})=1$,
$A=N=a^{\dagger}a$, $\left.|a_{n}\right\rangle =\left.|n\right\rangle $,
$\mu(d\mathbf{X})=d^{2}\alpha/\pi$, $\psi_{n}(\mathbf{X})=e^{-\frac{1}{2}|\alpha|^{2}}\dfrac{\left(\alpha^{*}\right)^{n}}{\sqrt{n!}}$,
with the reproducing kernel given by $K(\mathbf{\mathbf{X}{}^{\prime}},\mathbf{X})=\left\langle \alpha\right.|\left.\mathbf{\beta}\right\rangle =e^{\alpha^{*}\beta-\frac{1}{2}\left(|\alpha|^{2}+|\beta|^{2}\right)}$.
By using the kernel property (62), one has another kernel given by
$K(\alpha,\beta)=\left|\left\langle \alpha\right.|\left.\mathbf{\beta}\right\rangle \right|^{2}=e^{-|\alpha-\beta|^{2}}$
. 

A few examples of well known generalized coherent states that can
be derived using the above methodology are as follows. In \cite{gazeau2009coherent},
the authors use a 1+1 anti de-Sitter space constant negative curvature
metric $\mu(d\alpha,\gamma)={(2\gamma -1)d^2 \alpha}/\left[{\pi (1-|\alpha|^2)^2}\right]$ measure space and produce a reproducing kernel given by $(\alpha\in\mathbb{C},\gamma\in\mathbb{R})$
\begin{equation}
K(\mathbf{\mathbf{\alpha^{\prime}};\gamma},\mathbf{\alpha};\gamma)=\left(1-\left|\alpha\right|^{2}\right)^{\gamma}\left(1-\alpha^{\prime*}\alpha\right)^{-2\gamma}\left(1-\left|\alpha^{\prime}\right|^{2}\right)^{\gamma}.
\end{equation}
Generalized coherent states using a basis of Pöschl-Teller states
\cite{gazeau2009coherent} may be shown to produce a modified Bessel
function ($I_{_{\nu}}$) kernel given by

\begin{equation}
\begin{array}{c}
K(\mathbf{\mathbf{\alpha^{\prime}}},\mathbf{\alpha};\nu)={\Gamma}(\nu +1) \dfrac{I_{\nu}\left(2\left\{{|\alpha^{\prime}|^{2}}{|\alpha|^{2}}\right\} ^{1/4}\right)}
{\left\{{|\alpha^{\prime}|^{2}}{|\alpha|^{2}}\right\} ^{{\nu}/4}}\\
\\
\end{array}
\end{equation}
Several authors \cite{marhic1978oscillating,philbin2014generalized,senitzky1954harmonic}
have investigated oscillating Hermite polynomial Gaussian type wave
functions that lead to a reproducing kernel given by 
\begin{equation}
\begin{array}{c}
K(\alpha,\beta)=\\
\mathbf{exp}\left[-\left(\left|\alpha\right|^{2}+\left|\beta\right|^{2}-2\alpha^{*}\beta\right)/2\right]L_{n}(\left|\alpha-\beta\right|^{2})
\end{array}
\end{equation}
where $L_{n}$ are Laguerre polynomials. 

The nonlinear optical properties of confining Pöschl-Teller potentials
have been studied in \cite{yildirim2005nonlinear,csakirouglu2012nonlinear}.
Experimental realizations of Laguerre type generalized coherent states
using the behavior of a beamsplitter with a specific geometry are
explored in \cite{philbin2014generalized}. They also appear in the
realization of a quantum oscillator consisting of a trapped ion in
\cite{ziesel2013experimental}.The realization of generalized coherent
states have also been investigated in photonic lattices \cite{perez2011classical,perez2016generalized}.

For future work, each of these experiments needs to be analyzed in
a similar manner to that of section III in order to fully get an idea
of the quantum runtime efficiency. These recent experimental realizations
indicate that the quantum SVM approach to calculating reproducing
kernels from generalized coherent states may provide a fast and viable
alternative to classical computations.

\section{Conclusion}

In this paper, the potential of using generalized coherent states
as a calculational tool for quantum SVMs was demonstrated. The key
connecting thread was the RKHS concept used in SVMs. Such reproducing
kernels naturally arise in the quantum state overlap of canonical
and generalized coherent states. It was shown that canonical coherent
states reproduce the popular radial kernels of SVMs wherein POVM measurements
of overlap functions substantially reduce the computational times
of such kernels, especially in high dimensional feature spaces found
in big data sets. The use of reproducing kernels not usually used
in classical algorithms due to their complexity, such as those from
anti-de Sitter space coherent states and Bessel function kernels from
Pöschl-Teller coherent states are now conceivable using the coherent
state driven quantum SVM approach. The realization of generalized
coherent states via experiments in quantum optics indicate the near
term feasibility of this approach.
\begin{acknowledgments}
R.C. acknowledges partial support from the Chicago Mercentile Exchange
(CME) foundation.
\end{acknowledgments}


\end{document}